\documentclass[preprint,12pt]{elsarticle}

\usepackage[utf8]{inputenc}
\usepackage[T1]{fontenc}
\usepackage{amsmath,amssymb,amsfonts}
\usepackage{graphicx}
\usepackage{booktabs}
\usepackage{array}
\usepackage{caption}
\usepackage{setspace}
\usepackage[hidelinks]{hyperref}
\usepackage{xurl}
\usepackage{xcolor}
\usepackage{enumitem}
\usepackage{float} 

\journal{}

\newcommand{\Wsyn}{\mathbf{W}_{\mathrm{syn}}}
\newcommand{\Qsyn}{\mathbf{Q}_{\mathrm{syn}}}
\newcommand{\Lrecon}{\mathcal{L}_{\mathrm{recon}}}
\newcommand{\Lcluster}{\mathcal{L}_{\mathrm{cluster}}}
\newcommand{\Lkl}{\mathcal{L}_{\mathrm{KL}}}
\newcommand{\Locc}{\mathcal{L}_{\mathrm{occ}}}

\begin{document}

\begin{frontmatter}

\title{scDNM-VAE: Interpretable deep clustering of single-cell RNA-seq data with signed dendritic gating}

\author[1,2,3]{Melih A\u{g}raz\corref{cor1}}
\cortext[cor1]{Corresponding author}
\ead{melih_agraz@brown.edu}
\author[3]{Deniz Karap\i{}nar}
\author[4]{Aysel Top\c{s}ir}
\author[5]{Qianying Cao}
\author[3]{Erol E\u{g}rio\u{g}lu}
\author[1,2,6,7]{Gaurav Choudhary}
%\author[1,2,3]{Melih A\u{g}raz\corref{cor1}}
%\ead{melih_agraz@brown.edu}
%\author[3]{Deniz Karap\i{}nar}
%\ead{deniz.karapinar@giresun.edu.tr}
%\author[4]{Aysel Tops\i{}r}
%\ead{aysel.topsir@std.yildiz.edu.tr}
%\author[5]{Qianying Cao}
%\ead{qianying_cao@brown.edu}
%\author[3]{Erol E\u{g}rio\u{g}lu}
%\ead{erol.egrioglu@giresun.edu.tr}
%\author[1,2,6,7]{Gaurav Choudhary}
%\ead{gaurav_choudhary@brown.edu}

\address[1]{Department of Medicine, Brown University, Providence, RI 02903, USA}
\address[2]{Division of Cardiology, Brown University Health, Providence, RI 02903, USA}
\address[3]{Department of Data Science and Analytics, Faculty of Arts and Sciences, Giresun University, Giresun, T\"urkiye}
\address[4]{Department of Industrial Engineering, Faculty of Mechanical Engineering, Yildiz Technical University, Istanbul, T\"urkiye}
\address[5]{Department of Mechanical Engineering, Rochester Institute of Technology, Rochester, NY 14623, USA}
\address[6]{Alpert Medical School of Brown University and Brown Health Cardiovascular Institute, Providence, RI 02903, USA}
\address[7]{VA Providence Healthcare System, Providence, RI 02908, USA}

\begin{abstract}
Deep clustering models for single-cell RNA sequencing commonly assign cells through latent distances or centroid-based rules that are difficult to inspect directly. We introduce scDNM-VAE (single-cell Dendritic Neuron Model Variational Autoencoder), which couples a variational encoder to a signed dendritic clustering head. Each cluster is defined by learnable weight--threshold gates: the weight sign determines whether a gate increases or decreases along a latent coordinate, its magnitude controls transition steepness, and the weight--threshold ratio determines the transition midpoint. The complete assignment rule can therefore be examined without fitting a post-hoc surrogate.

We evaluate scDNM-VAE on four public datasets spanning immune, cortical, cardiac, and hematopoietic cells, using five matched random seeds and locally rerun MLP-DEC and scVI+KMeans comparators. An unsupervised occupancy-prior term discourages drift from the populated KMeans initialization, while a mild sharpness schedule gradually introduces the final gate steepness. Among the three primary benchmark methods, scDNM-VAE achieves the
highest mean adjusted Rand index on PBMC3k (0.783) and Heart (0.758),
is within 0.009 of scVI+KMeans on Zeisel (0.687 versus 0.696), and
trails MLP-DEC on Paul15 (0.298 versus 0.319). A matched head-only ablation shows dataset-dependent effects: the
dendritic head yields higher mean ARI than a Student-$t$ DEC head on
PBMC3k and Zeisel, whereas the DEC head performs better on Heart and
Paul15. These findings support signed dendritic gating as
a directly inspectable alternative for cluster assignment, rather than
a universally accuracy-enhancing replacement for centroid-based rules.
\end{abstract}


\end{frontmatter}

\section{Introduction}

Single-cell RNA sequencing (scRNA-seq) profiles the transcriptomes of individual cells and resolves cell types, transitional states, and rare populations that bulk assays miss. Unsupervised clustering of these expression profiles is one of the first steps in nearly every scRNA-seq pipeline, and the quality of that clustering shapes downstream analyses and interpretation of differential expression, trajectory inference, and cell--cell communication~\cite{1,5,6}.

In deep single-cell analysis, predictive or clustering performance alone does not reveal how a model reaches an assignment. Post-hoc methods such as SHAP and LIME can summarize model behaviour, but they approximate a trained decision function rather than expose it directly~\cite{35,40,41,42}. This distinction matters because clustering choices propagate into differential-expression, trajectory, and cell--cell communication analyses. An architecture whose assignment rule is directly inspectable can therefore complement conventional biological validation without claiming that parameter readability alone establishes biological correctness.

Current deep-learning approaches to scRNA-seq clustering commonly follow one of two designs. Two-stage pipelines combine representation-learning models such as scVI~\cite{1} with algorithms such as KMeans or Leiden~\cite{10}; joint methods, including DEC~\cite{4} and scDeepCluster~\cite{5}, optimize representation and clustering together. Their assignments are typically expressed through centroid distances or target-distribution refinement. These rules are mathematically explicit, but their learned parameters do not directly state whether increasing a particular latent coordinate promotes or suppresses membership in a cluster. Biological interpretation is therefore usually added downstream through differential expression, marker enrichment, or another explanatory analysis.
Supervised cell-type annotation methods and benchmarking studies have further demonstrated the importance of robust cell-identity assignment in scRNA-seq analysis, although these approaches generally assume labelled reference data rather than solving the unsupervised clustering problem directly~\cite{22,23,26,32}. Semi-supervised approaches have also been explored in clinical
prediction tasks, including thyroid disease classification using
pseudo-labeling and FixMatch-based learning~\cite{48}.
Unlike these approaches, scDNM-VAE addresses unsupervised
cell clustering without using reference cell-type labels
during training. Autoencoder-based denoising methods such as DCA have also shown the utility of deep autoencoder-based representations for handling sparse and noisy scRNA-seq count data~\cite{27}. Related work has also pursued explainable neural-network models for
single-cell analysis; for example, CellTICS combines cell-type identification
with gene-level interpretation from scRNA-seq data~\cite{25}. Related studies have also explored representation learning
and feature selection for genomic data analysis. For example,
ML-GAP combines autoencoder-based representation learning,
data augmentation, and feature selection to support
machine-learning analysis of RNA-seq data~\cite{45}. Feature selection is particularly relevant to high-dimensional
gene-expression analysis, where identifying informative molecular
features can facilitate downstream modeling. Previous comparisons
of LASSO, Boruta, and minimum-redundancy--maximum-relevance
methods in breast cancer microarray data illustrate this
methodological challenge~\cite{46}.

In contrast, the architecture of the dendritic neuron model (DNM) introduced by Todo and colleagues~\cite{2} has properties that support direct parameter-level inspection. DNM was designed to capture how biological neurons integrate signals along their dendritic trees: each synapse applies a nonlinear sigmoid gate parameterized by a weight and a threshold (Figure~\ref{fig:dnm}a), the synaptic outputs along each dendritic branch are combined multiplicatively, and the resulting branch outputs are then summed at the membrane/soma layer~\cite{3,13}. It has been used for classification~\cite{14,15,20,30}, function approximation~\cite{3}, and forecasting~\cite{28,31}, with accuracy comparable to standard neural networks but with parameters that remain inspectable~\cite{16,17}. A range of DNM variants and training strategies have since been proposed, including models with explicit dendrite morphology, adaptive synapses, and multi-input/multi-output extensions~\cite{18,19,29}. Related work on non-standard neuron architectures has also explored multiplicative aggregation, including single multiplicative neuron models within physics-informed neural networks, where architectural simplicity offered computational advantages in some settings but also revealed important convergence limitations~\cite{44}. Two properties make this architecture particularly appealing for clustering. The threshold-gated formulation carves the latent space into regions defined by distinct activation patterns, which is a stronger inductive bias than Euclidean distance to a centroid; note that our Stage-1 head aggregates the per-dimension gates additively, as $\mathrm{dist}_k=\sum_d (1-A_{kd})^2$, and reserves the DNM's multiplicative branch aggregation for the Stage-2 gene-module decoder. Crucially, each signed weight-threshold pair specifies how a cluster reads one latent coordinate: the weight sign determines whether the gate increases or decreases, the magnitude controls the steepness of this dependence, and the weight--threshold pair determines the transition location. These quantities are part of the clustering decision function itself and can therefore be inspected without fitting a separate post-hoc model. In scDNM-VAE, we adapt the DNM principle differently across the two stages: Stage~1 uses signed weight-threshold gating for cluster assignment, whereas Stage~2 uses a modified dendritic branch aggregation based on the geometric mean for gene-module decomposition.

A growing body of work pursues \emph{inherent} interpretability in single-cell models by constraining the latent space to align with known biology. Biologically informed VAEs such as expiMap~\cite{36} and VEGA~\cite{37} tie individual latent dimensions to curated gene programs or pathways, while linearly-decoded VAEs (LDVAE~\cite{38}) and embedded topic models (scETM~\cite{39}) expose interpretable gene loadings directly from the decoder. These methods make the \emph{gene-program content} of the latent space readable. Our approach is complementary but operates on a different axis: rather than annotating what each latent dimension represents, scDNM-VAE exposes the signed weight-threshold gates that determine how each cluster reads the latent space, and we test by ablation whether the dimensions receiving the largest synaptic weights influence the resulting assignments.

In this study, we report scDNM-VAE, a pipeline designed from the start to make its clustering rule directly inspectable, building on the DNM~\cite{2} for clustering cells from scRNA-seq data. We evaluate it against scVI+KMeans and an MLP-DEC ablation on four datasets covering immune cells, cortical neurons, cardiac tissue, and hematopoietic progenitors. Across five matched random seeds, scDNM-VAE has the highest mean ARI
among the three primary benchmark methods on PBMC3k and Heart, performs
within 0.009 ARI of scVI+KMeans on Zeisel, and is below MLP-DEC on
Paul15. A separate matched head-only ablation shows that replacing the
dendritic head with a Student-$t$ DEC head improves performance on Heart
and Paul15 but reduces it on PBMC3k and Zeisel. We therefore view the main contribution as an inspectable clustering rule that maintains competitive performance, rather than a uniformly more accurate clustering method. We further evaluate whether conventional cluster-quality metrics faithfully reflect biological structure and examine the extent to which the largest signed gating parameters are decision-relevant. Distinct from a post-hoc explanation, the signed gates are components of the clustering rule itself and can be read directly from the trained model. The following sections present the model architecture, benchmarking results, biological characterization, and interpretability analyses, while the Methods section details the datasets, preprocessing procedures, model implementation, comparator configurations, ablation protocols, and evaluation framework.

\section{Results}

\begin{figure}[H]
\centering
\includegraphics[width=\textwidth]{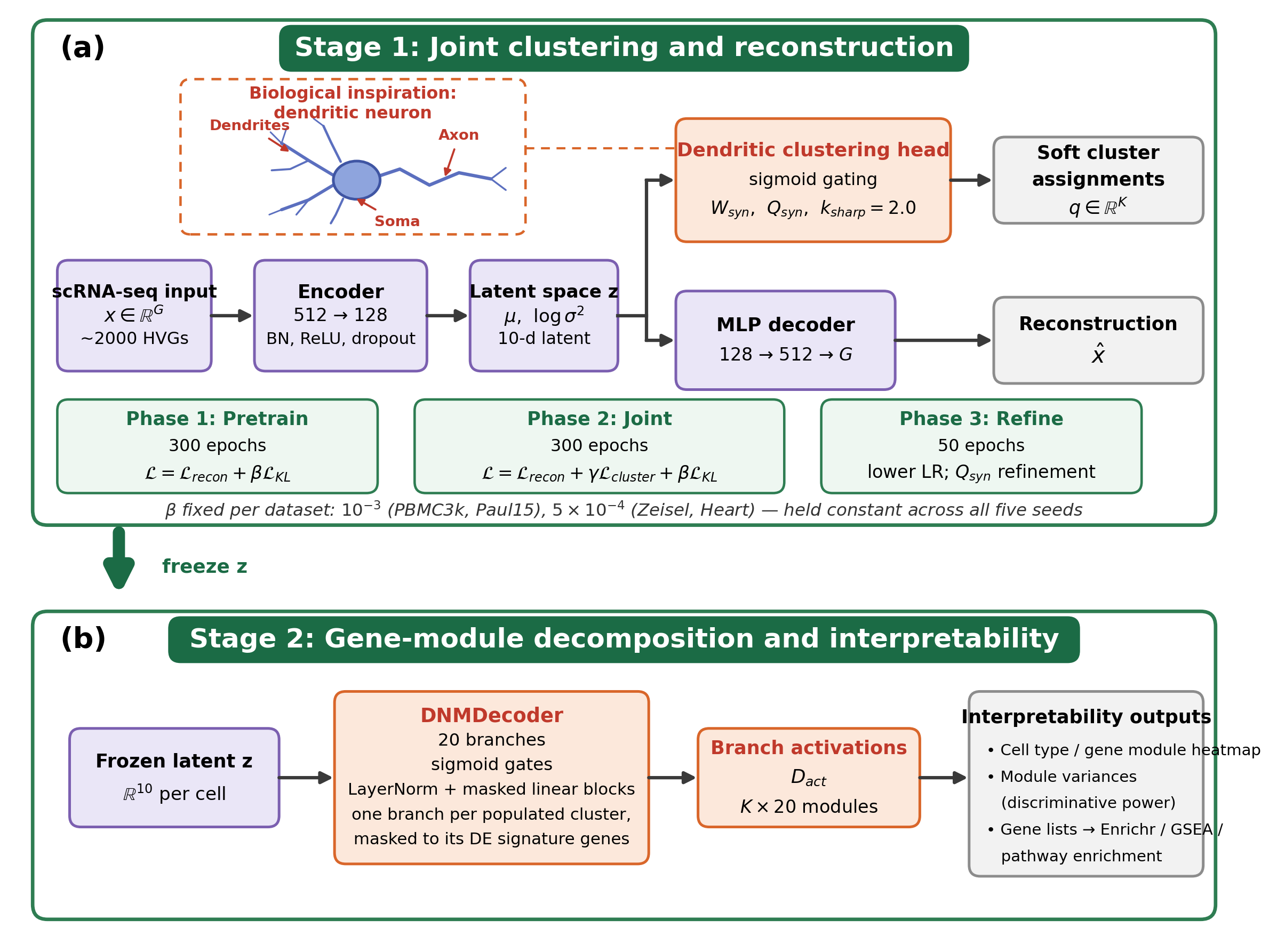}
\caption{\textbf{scDNM-VAE training and interpretation workflow.} (a)~Stage~1 learns a 10-dimensional variational representation and cluster assignments through signed weight--threshold gates. Posterior means from pretraining initialize KMeans centres, which initialize $\Wsyn$ and $\Qsyn$ and define the unsupervised occupancy prior. Phase~1 pretrains reconstruction and KL terms; Phase~2 jointly optimizes reconstruction, DEC clustering, variational KL, and occupancy regularization while $k_{\mathrm{sharp}}$ increases from 1.5 to 2.0 over the first 75 epochs; Phase~3 recomputes $\Qsyn$ from the optimized $\Wsyn$ and briefly refines the model at $k_{\mathrm{sharp}}=2.0$. (b)~Stage~2 freezes the Stage-1 latent representation and predicted assignments, constructs cluster-aligned differential-expression masks, and trains a 20-branch DNMDecoder to obtain branch-activity summaries and gene modules. Ground-truth labels are used only for external evaluation, and Stage~2 does not alter the Stage-1 clustering solution.}
\label{fig:arch}
\end{figure}

\subsection{Model architecture}

scDNM-VAE operates in two stages (Figure \ref{fig:arch}). Stage~1 consists of a shared variational encoder, a DNM-inspired dendritic clustering head, and an MLP decoder, whereas Stage~2 augments the model with a dendritic gene-module decoder trained on frozen Stage 1 latent representations that provides complementary gene-level interpretability without altering the Stage~1 clustering solution. Importantly, the Stage~1 clustering head uses synaptic weight-threshold gating to define cluster assignments in the latent space, while the geometric-mean branch aggregation is used in the Stage~2 dendritic decoder to decompose latent representations into gene modules. The encoder maps a cell's expression profile $\mathbf{x}\in\mathbb{R}^{G}$ through two hidden layers (512 and 128 units; BatchNorm, ReLU, 10\% dropout) to produce a mean vector $\boldsymbol{\mu}$ and a log-variance vector $\log\boldsymbol{\sigma}^{2}$ in a 10-dimensional latent space. A latent sample $\mathbf{z}$ is drawn via the reparameterization trick, $\mathbf{z}=\boldsymbol{\mu}+\boldsymbol{\varepsilon}\cdot\boldsymbol{\sigma}$, where $\boldsymbol{\varepsilon}\sim\mathcal{N}(\mathbf{0},\mathbf{I})$~\cite{11}.

\begin{figure}[H]
\centering
\includegraphics[width=\textwidth]{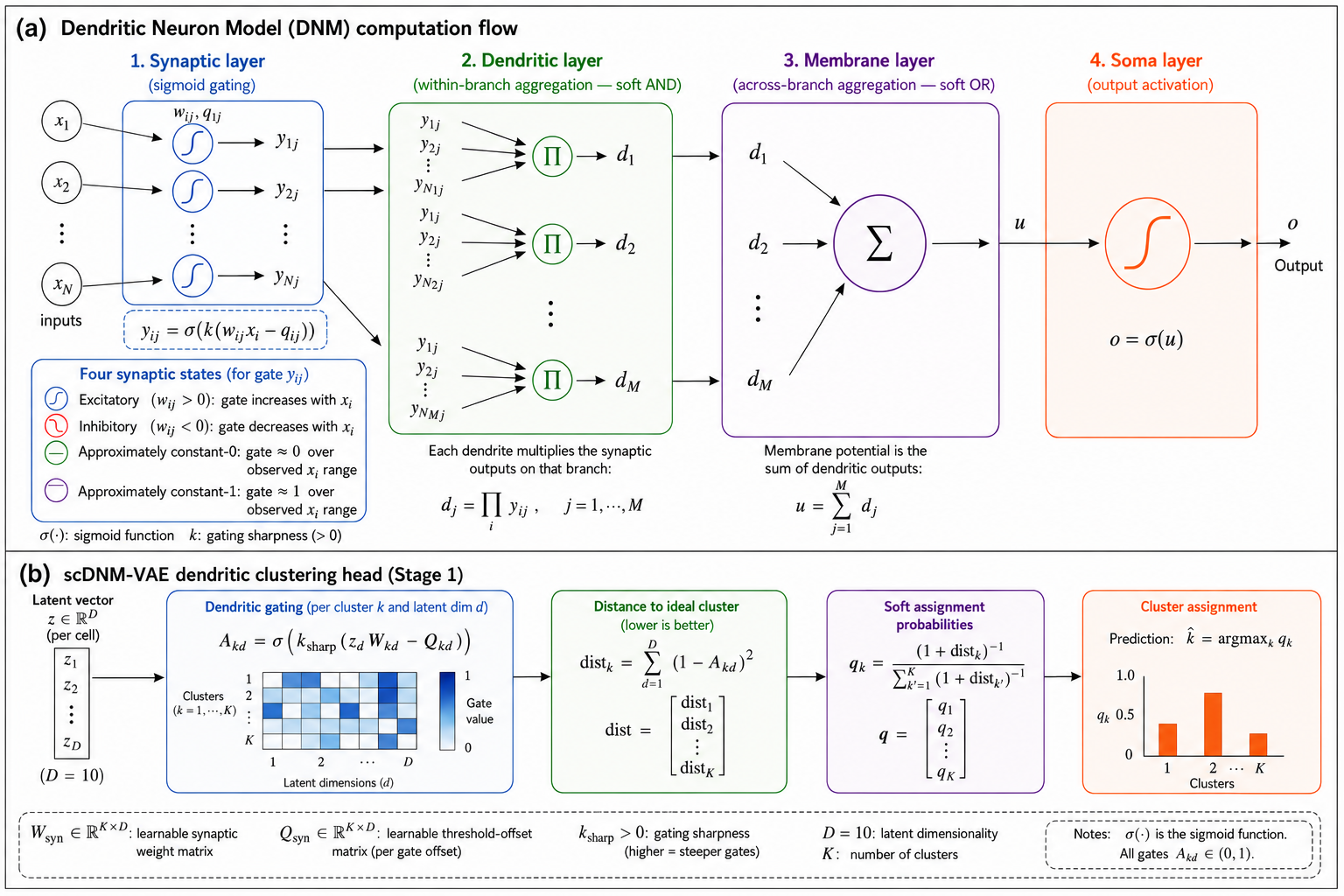}
\caption{\textbf{The dendritic neuron model and the scDNM-VAE Stage-1
clustering head.}
(a)~Computation flow of the classical DNM~\cite{2,3}. Each input $x_i$ reaches
each branch $j$ through its own synapse, gated by
$y_{ij}=\sigma\!\big(k(w_{ij}x_i-q_{ij})\big)$, where $k>0$ sets the gating
sharpness. Gated outputs are combined multiplicatively within a branch
($d_j=\prod_i y_{ij}$), summed across branches ($u=\sum_j d_j$), and passed
through a final sigmoid. The four synaptic regimes are defined jointly by
$w_{ij}$ and $q_{ij}$ relative to the observed range of $x_i$, not by the sign
of $w_{ij}$ alone.
(b)~The Stage-1 clustering head. Each cluster $k$ reads each latent dimension
$d$ through a signed gate
$A_{kd}=\sigma\!\big(k_{\mathrm{sharp}}(z_dW_{kd}-Q_{kd})\big)$, with
$k_{\mathrm{sharp}}$ annealed from 1.5 to 2.0 during the first quarter of joint training and thresholds capped at $Q_{\max}=1.5$. Gate openings
are aggregated into $\mathrm{dist}_k=\sum_d(1-A_{kd})^2$, converted to soft
assignments $q_k$, and each cell is assigned to $\arg\max_k q_k$. The sign of
$W_{kd}$ determines the direction of the gate, $|W_{kd}|$ its steepness, and
$Q_{kd}/W_{kd}$ the latent value at which it switches; these three quantities
are components of the clustering rule itself and are read directly from the
trained parameters, without fitting a separate post-hoc explanation model.
Stage~1 aggregates the per-dimension gates \emph{additively}; the multiplicative
branch aggregation of panel~(a) is used only by the Stage-2 gene-module decoder
(Methods, Complementary gene-module decoder). Because $A_{kd}\in(0,1)$, $\mathrm{dist}_k$ lies in $(0,D)$ and never
reaches zero; it is therefore an assignment score rather than a metric
distance, and $q_k$ is correspondingly bounded away from~1.}
\label{fig:dnm}
\end{figure}

The dendritic clustering head implements the DNM gating mechanism. For each cluster $k$ and latent dimension $d$, a sigmoid gate is computed as

\[
A_{kd} = \sigma\!\big(k_{\mathrm{sharp}}\,(z_d\,W_{kd}-Q_{kd})\big),
\]
 
where $\Wsyn$ and $\Qsyn$ are learnable synaptic weight and threshold matrices. During joint training, $k_{\mathrm{sharp}}$ increases linearly from 1.5 to 2.0 over the first quarter of Phase~2 and remains at 2.0 thereafter. This mild continuation schedule avoids imposing the final gate steepness at the start of joint optimization. The gate-based assignment score for cluster $k$ is
$\mathrm{dist}_k=\sum_d (1-A_{kd})^2$, which decreases as that cluster's
gates open, and the soft assignment probability is
$q_k=(1+\mathrm{dist}_k)^{-1}$, normalized across clusters
(Figure~\ref{fig:dnm}). We retain the symbol $\mathrm{dist}_k$ for
continuity with the DEC formulation, but it is not a metric: it is bounded,
never reaches zero, and is minimized not at a centroid but along the
direction of $\mathbf{W}_k$. The gate derivative is
$\partial A_{kd}/\partial z_d
=k_{\mathrm{sharp}}W_{kd}A_{kd}(1-A_{kd})$,
so the sign of $W_{kd}$ determines whether the gate increases or
decreases with the latent coordinate, while $|W_{kd}|$ controls the
steepness and, for $W_{kd}\neq0$, the transition midpoint occurs at
$z_d=Q_{kd}/W_{kd}$. The decoder reconstructs the input through a symmetric MLP that mirrors the encoder (128 and 512 hidden units), producing a $G$-dimensional output matching the input gene space.

Training proceeds in three phases. Phase~1 (300 epochs) pretrains the encoder and decoder using reconstruction and KL-divergence losses. In Phase~2 (300 epochs), KMeans is fitted to the pretrained posterior-mean representations, and its cluster centres are copied into the rows of $\Wsyn$; $\Qsyn$ is then initialized from the nearest-neighbour distances among these rows before joint optimization with the DEC-style target distribution. The mean soft KMeans assignment defines an unsupervised occupancy prior, and a KL penalty discourages the learned mean assignment from collapsing away from that populated initialization. Gate sharpness is annealed from 1.5 to 2.0 over the first 75 Phase~2 epochs. In Phase~3 (50 epochs), $\Qsyn$ is recomputed from the optimized $\Wsyn$ using the same distance-based rule, followed by brief refinement at the final sharpness and a lower learning rate.

\paragraph{KL regularization}
For the reported scDNM-VAE experiments, the KL coefficient was set to
$\beta=10^{-3}$ for PBMC3k and Paul15 and
$\beta=5\times10^{-4}$ for Zeisel and Heart. Within each dataset,
the same value was held fixed across all five random seeds.

\subsection{Benchmark performance}

We benchmarked scDNM-VAE, the MLP-DEC ablation, and scVI+KMeans on four datasets with expert-curated labels (Tables~\ref{tab:ari} and~\ref{tab:bench}). All three methods were rerun locally with the same five random seeds (42, 7, 123, 256, 999). Mean ARI across seeds is the primary summary because it is less sensitive than selecting the single best run; best-seed ARI is reported descriptively in Table~\ref{tab:bench}. scVI received retained raw counts for its count likelihood, whereas scDNM-VAE and MLP-DEC received the same normalized, highly variable-gene input matrix. The cluster number was fixed to the number of reference labels for every method.

Among the three primary benchmark methods, scDNM-VAE achieves the
highest mean ARI on two of four datasets (Table~\ref{tab:ari}). On PBMC3k, its mean ARI is 0.783, compared with 0.733 for MLP-DEC and 0.463 for scVI+KMeans. On Heart, scDNM-VAE reaches 0.758, compared with 0.717 and 0.716, respectively. These comparisons support a consistent advantage on PBMC3k and a smaller advantage on Heart under the matched local protocol.

Zeisel represents a near tie in mean performance: scVI+KMeans reaches 0.696 and scDNM-VAE 0.687, a difference of 0.009 that is smaller than the scDNM-VAE seed standard deviation (0.064). scDNM-VAE nevertheless attains the highest individual run on this dataset (ARI 0.797), illustrating why we use the five-seed mean, rather than best-seed selection, as the primary result. On Paul15, MLP-DEC has the highest mean ARI (0.319), followed by scDNM-VAE (0.298) and scVI+KMeans (0.273). Paul15 therefore remains a clear limitation rather than evidence of a universal gain.

Across the three primary benchmark methods, scDNM-VAE ranks first twice,
second twice, and never third by mean ARI. Relative to MLP-DEC, its mean
ARI is higher on PBMC3k, Zeisel, and Heart but lower on Paul15. Because
MLP-DEC also differs in its deterministic encoding scheme, KL
regularization, occupancy regularization, and training schedule, that
comparison alone cannot isolate the effect of dendritic gating. We
therefore performed a separate matched head-only VAE--DEC ablation, reported
below.

\begin{table}[t]
\centering
\caption{\textbf{Primary benchmark: ARI mean $\pm$ sample standard deviation across five matched random seeds.} All methods were trained and evaluated locally under their method-appropriate input models; bold indicates the largest mean within a row.}
\label{tab:ari}
\small
\begin{tabular}{lccc}
\toprule
\textbf{Dataset} & \textbf{scDNM-VAE} & \textbf{MLP-DEC} & \textbf{scVI+KMeans} \\
\midrule
PBMC3k & \textbf{0.783 $\pm$ 0.019} & 0.733 $\pm$ 0.029 & 0.463 $\pm$ 0.018 \\
Zeisel & 0.687 $\pm$ 0.064 & 0.474 $\pm$ 0.010 & \textbf{0.696 $\pm$ 0.015} \\
Heart & \textbf{0.758 $\pm$ 0.047} & 0.717 $\pm$ 0.005 & 0.716 $\pm$ 0.084 \\
Paul15 & 0.298 $\pm$ 0.012 & \textbf{0.319 $\pm$ 0.012} & 0.273 $\pm$ 0.009 \\
\bottomrule
\end{tabular}
\end{table}

\begin{figure}[H]
\centering
\includegraphics[width=\textwidth]{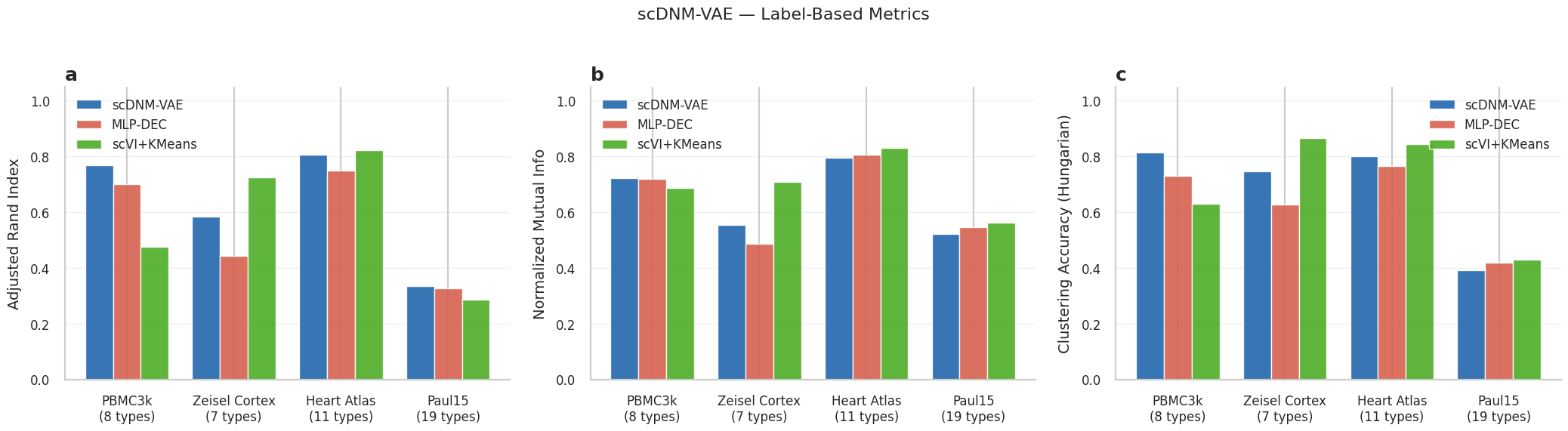}
\caption{\textbf{Label-based clustering performance across five matched random seeds.} (a)~Mean adjusted Rand index (ARI) and (b)~mean normalized mutual information (NMI) for scDNM-VAE (blue), MLP-DEC (red), and scVI+KMeans (green). Error bars denote one sample standard deviation. ARI values correspond to Table~\ref{tab:ari}.}
\label{fig:metrics}
\end{figure}

\begin{table}[t]
\centering
\caption{\textbf{Best-seed ARI across four scRNA-seq datasets.} These values describe the highest individual run among five matched seeds and are secondary to the mean $\pm$ standard deviation in Table~\ref{tab:ari}. Bold indicates the largest value within a row. $\Delta$ vs scVI is calculated from the displayed rounded values.}
\label{tab:bench}
\small
\resizebox{\linewidth}{!}{%
\begin{tabular}{lrrrcccr}
\toprule
\textbf{Dataset} & \textbf{Cells} & \textbf{Types} & $\boldsymbol{\beta}$ & \textbf{scDNM-VAE} & \textbf{MLP-DEC} & \textbf{scVI+KM} & \textbf{$\Delta$ vs scVI} \\
\midrule
PBMC3k & 2{,}638 & 8 & 0.001 & \textbf{0.804} & 0.784 & 0.490 & $+0.314$ \\
Zeisel & 3{,}005 & 7 & 0.0005 & \textbf{0.797} & 0.488 & 0.712 & $+0.085$ \\
Heart & 18{,}641 & 11 & 0.0005 & \textbf{0.829} & 0.723 & 0.811 & $+0.018$ \\
Paul15 & 2{,}730 & 19 & 0.001 & 0.317 & \textbf{0.337} & 0.287 & $+0.030$ \\
\bottomrule
\end{tabular}%
}
\end{table}
The corresponding ARI and NMI results across the five matched random seeds are shown in Figure~\ref{fig:metrics}, which provides a visual summary of the between-method performance differences and seed-to-seed variability.

Figure~\ref{fig:metrics} shows that the relative performance patterns are
generally consistent across ARI and NMI, although the ordering of methods
is not identical for every dataset. scDNM-VAE shows the clearest advantage
on PBMC3k, where it attains the highest mean values for both ARI and NMI.
On Zeisel, scVI+KMeans has a slightly higher mean ARI, whereas the NMI
values of scDNM-VAE and scVI+KMeans are nearly identical, with scDNM-VAE
showing substantially greater seed-to-seed variability. On Heart, scDNM-VAE
has the highest mean ARI, while scVI+KMeans achieves the highest mean NMI;
the latter also shows comparatively large variability across seeds. Paul15
remains the most difficult benchmark overall, and MLP-DEC achieves the
highest mean values for both ARI and NMI on this dataset. These results
indicate that the relative ranking of the methods depends not only on the
dataset but also, to some extent, on the agreement metric considered.
\subsection{Geometric tightness does not imply biological correctness}

Geometry-based metrics were computed from the dataset-specific best-ARI runs used for the benchmark (Supplementary Figure~S1). Across all four datasets, MLP-DEC produced geometrically tighter partitions than scDNM-VAE, with higher silhouette and Calinski--Harabasz scores and lower Davies--Bouldin scores. This ordering did not track agreement with the reference cell types: scDNM-VAE achieved the higher best-seed ARI on three of the four datasets, whereas Paul15 was the exception (Table~\ref{tab:bench}). On the Heart Atlas, MLP-DEC's Calinski--Harabasz score was particularly large despite its lower ARI. These results show that compactness in a learned latent space and correspondence with biological reference labels measure different properties; neither should be treated as a surrogate for the other.

\paragraph{Matched head-only VAE--DEC ablation}
To isolate the contribution of the clustering head more closely, we
replaced the signed dendritic assignment function with a Student-$t$ DEC
centroid head while retaining the stochastic VAE encoder, decoder,
10-dimensional latent representation, dataset-specific KL coefficient,
KMeans-derived occupancy prior, preprocessing, initialization, and matched
training duration. Across five seeds, scDNM-VAE achieved higher mean ARI
on PBMC3k (0.783 versus 0.549) and Zeisel (0.687 versus 0.649), whereas
the matched VAE--DEC variant performed better on Heart (0.820 versus
0.758) and Paul15 (0.340 versus 0.298). These results suggest that the benefit of the dendritic head depends on the dataset
(Table~\ref{tab:headonly}).

\begin{table}[t]
\centering
\caption{\textbf{Matched head-only ablation: ARI mean $\pm$ sample
standard deviation across five random seeds.} The stochastic VAE encoder,
decoder, latent dimensionality, KL regularization, KMeans-derived occupancy
prior, preprocessing, initialization, and training duration were matched;
only the dendritic assignment head was replaced by a Student-$t$ DEC
centroid head. Bold indicates the larger mean within a row.}
\label{tab:headonly}
\small
\begin{tabular}{lcc}
\toprule
\textbf{Dataset} & \textbf{scDNM-VAE} &
\textbf{Head-only VAE--DEC} \\
\midrule
PBMC3k & \textbf{0.783 $\pm$ 0.019} & 0.549 $\pm$ 0.012 \\
Zeisel & \textbf{0.687 $\pm$ 0.064} & 0.649 $\pm$ 0.083 \\
Heart  & 0.758 $\pm$ 0.047 & \textbf{0.820 $\pm$ 0.039} \\
Paul15 & 0.298 $\pm$ 0.012 & \textbf{0.340 $\pm$ 0.010} \\
\bottomrule
\end{tabular}
\end{table}

\begin{figure}[H]
\centering
\includegraphics[width=\textwidth]{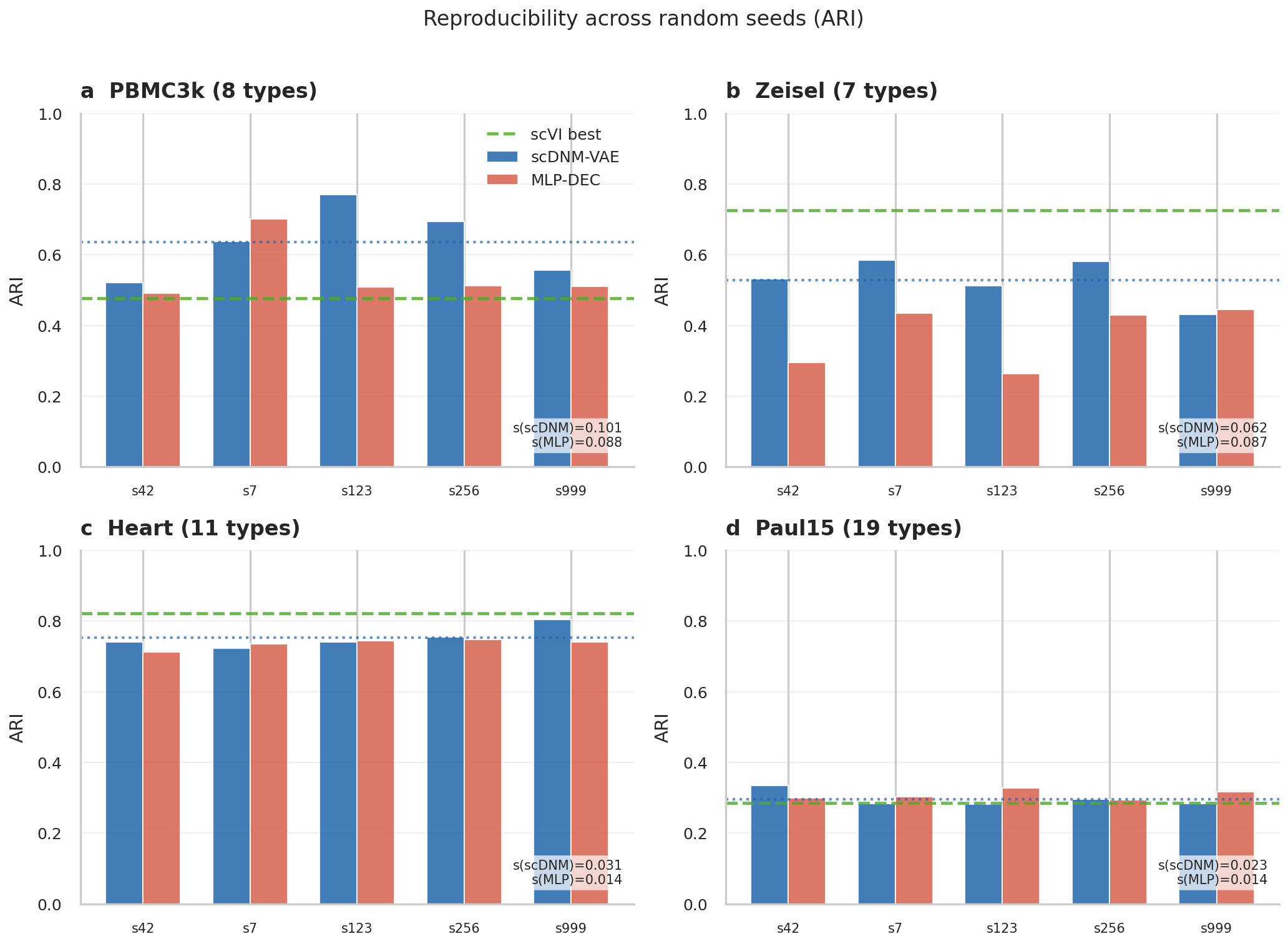}
\caption{\textbf{Seed-to-seed stability of clustering performance.} Per-seed ARI for scDNM-VAE (blue circles), MLP-DEC (red squares), and scVI+KMeans (green triangles) on PBMC3k, Zeisel, Heart, and Paul15 using the same five seeds (42, 7, 123, 256, 999). Solid horizontal lines in the corresponding method colors denote five-seed means. The plotted values are the individual runs summarized as mean $\pm$ sample standard deviation in Table~\ref{tab:ari}.}
\label{fig:seeds}
\end{figure}

\paragraph{Seed-to-seed stability}
Figure~\ref{fig:seeds} shows the five matched-seed ARI values underlying Table~\ref{tab:ari}. Stability is dataset dependent. On PBMC3k, scDNM-VAE combines the highest mean with low variability ($0.783\pm0.019$). Its variability is greatest on Zeisel ($\sigma=0.064$), where one high-performing seed raises the best-run ARI but the five-seed mean remains slightly below scVI+KMeans. On Heart, scVI+KMeans is the most variable method ($\sigma=0.084$), whereas MLP-DEC is highly stable but has a lower mean than scDNM-VAE. All three methods show comparatively small variation on Paul15 ($\sigma=0.009$--$0.012$), indicating that the lower ARI values there are systematic under the evaluated configurations rather than driven by a single unfavorable seed.

\subsection{Biological characterization through differential expression}

To check whether the resulting clusters carry biological meaning, we ran
Wilcoxon rank-sum differential-expression analysis with
Scanpy~\cite{6}. Top marker genes recovered recognizable cell-type
signatures across all four datasets. In PBMC3k, the CD14$^+$ monocyte
cluster was characterized by \emph{LYZ}, \emph{TYROBP}, \emph{LST1},
and \emph{S100A9}, whereas the B-cell cluster was characterized by
\emph{CD79A}, \emph{CD74}, and \emph{CD79B}. In the Human Heart Cell
Atlas, ventricular and atrial cardiomyocyte clusters were marked by
\emph{RYR2}/\emph{FHL2} and \emph{MYL7}/\emph{MYH6}/\emph{NPPA},
respectively; endothelial, fibroblast, macrophage, and pericyte clusters
included \emph{TM4SF1}, \emph{DCN}, \emph{F13A1}, and \emph{PDGFRB},
respectively. In Zeisel Cortex, oligodendrocyte and interneuron clusters
included \emph{Mobp} and \emph{Gad1}/\emph{Gad2}, respectively. In
Paul15, erythroid and granulocytic populations included
\emph{Car1}/\emph{Car2}/\emph{Ermap} and
\emph{Mpo}/\emph{Elane}/\emph{Ctsg}, respectively. The displayed genes
are representative members of the cluster-specific lists reported in
Supplementary Table~S1; the separate curated panels used for automatic
marker-overlap annotation are provided in Supplementary Table~S2.

\begin{figure}[p]
\centering
\includegraphics[width=\textwidth]{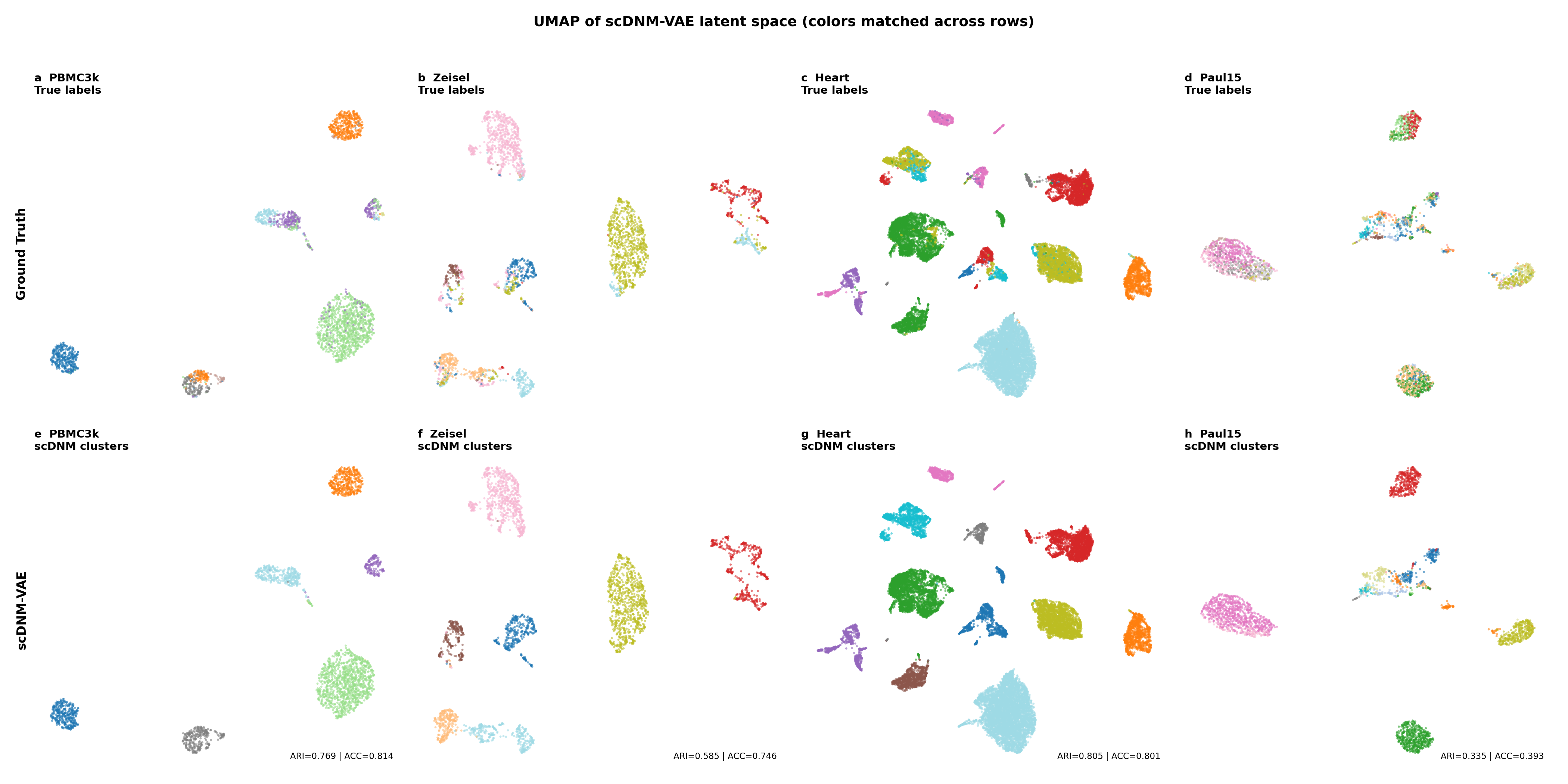}
\caption{\textbf{UMAP visualization of the scDNM-VAE latent space.} Each row presents one benchmark dataset: PBMC3k (a,b), Zeisel Cortex (c,d), Human Heart Cell Atlas (e,f), and Paul15 (g,h). The left panel of each pair colours cells by the reference cell-type labels used for evaluation. The right panel shows the identical two-dimensional coordinates partitioned by the scDNM-VAE predictions, with cluster identifiers placed at their median UMAP locations. For visual correspondence only, each predicted cluster is coloured post hoc according to its dominant reference cell type; reference labels were not used to train the model or determine cluster assignments. Consequently, clusters with the same dominant reference type share a colour but remain distinguishable by their cluster identifiers. UMAP was fitted to the saved 10-dimensional Stage-1 latent representation using 15 neighbours, a minimum distance of 0.25, Euclidean distance, and random seed 42. The displayed latent representations are the dataset-specific best-ARI runs from the five-seed benchmark (PBMC3k seed 256, Zeisel seed 123, Heart seed 42, and Paul15 seed 42); no comparator-method embedding is shown.}
\label{fig:umap}
\end{figure}

The automatic marker-overlap annotation (Methods, Cluster annotation by
marker overlap) agreed exactly with the dominant harmonized reference label
for all seven Zeisel clusters, nine of ten eligible Heart clusters, six of
eight PBMC3k clusters, and six of sixteen eligible Paul15 clusters
(Table~\ref{tab:annot}). These figures were computed from the same
dataset-specific best-seed runs used in Figures~\ref{fig:umap} and
\ref{fig:heartconf}. They are conditional panel--cluster agreement rates,
not independent measures of clustering accuracy. The contrast between the
strong agreement on Zeisel and Heart and the substantially weaker agreement
on Paul15 is consistent with the dataset-level ARI results. Visual inspection of the paired latent-space
Uniform Manifold Approximation and Projection
(UMAP) panels (Figure~\ref{fig:umap}) likewise shows strong cluster--label
correspondence on Zeisel and Heart, whereas Paul15 exhibits substantially
more extensive cell-type mixing.

\begin{table}[t]
\centering
\caption{
\textbf{Harmonized panel-category-level agreement of the automatic
marker-panel annotation with dominant reference labels}
(Methods, Cluster annotation by marker overlap). Reference labels were first mapped onto the curated
marker-panel vocabulary (Supplementary Table~S3), which merges some finer
reference categories into broader panel categories, for example atrial and
ventricular cardiomyocytes into ``Cardiomyocytes.'' Both the dominant label
and the $\geq40\%$ eligibility criterion were computed after this mapping;
the reported agreement therefore applies at the harmonized panel-category
level rather than at the resolution of the original reference annotations.
``Eligible clusters'' is the number of predicted clusters whose dominant
harmonized reference label accounts for at least 40\% of the cluster's cells.
``Exact match'' counts eligible clusters whose automatic marker-overlap
annotation matches that dominant harmonized label, and ``Agreement'' is the
exact-match count divided by the number of eligible clusters. This measure
evaluates panel--cluster agreement, not clustering quality; direct
cluster--reference correspondence is shown by the per-cluster purity of the
Heart confusion matrix (Figure~\ref{fig:heartconf}).
Eligible clusters are not the total number of reference cell types.}
\label{tab:annot}
\small
\begin{tabular}{lccc}
\toprule
\textbf{Dataset} & \textbf{Eligible clusters ($\ge$40\%)} & \textbf{Exact match} & \textbf{Agreement} \\
\midrule
Zeisel Cortex     & 7  & 7 & 100.0\% \\
Heart Cell Atlas  & 10 & 9 & 90.0\% \\
PBMC3k            & 8  & 6 & 75.0\% \\
Paul15            & 16 & 6 & 37.5\% \\
\bottomrule
\end{tabular}
\end{table}

\subsection{Cluster-to-ground-truth correspondence on the Human Heart Cell Atlas} \label{sec:heart}

\begin{figure}[p]
\centering
\includegraphics[width=\textwidth]{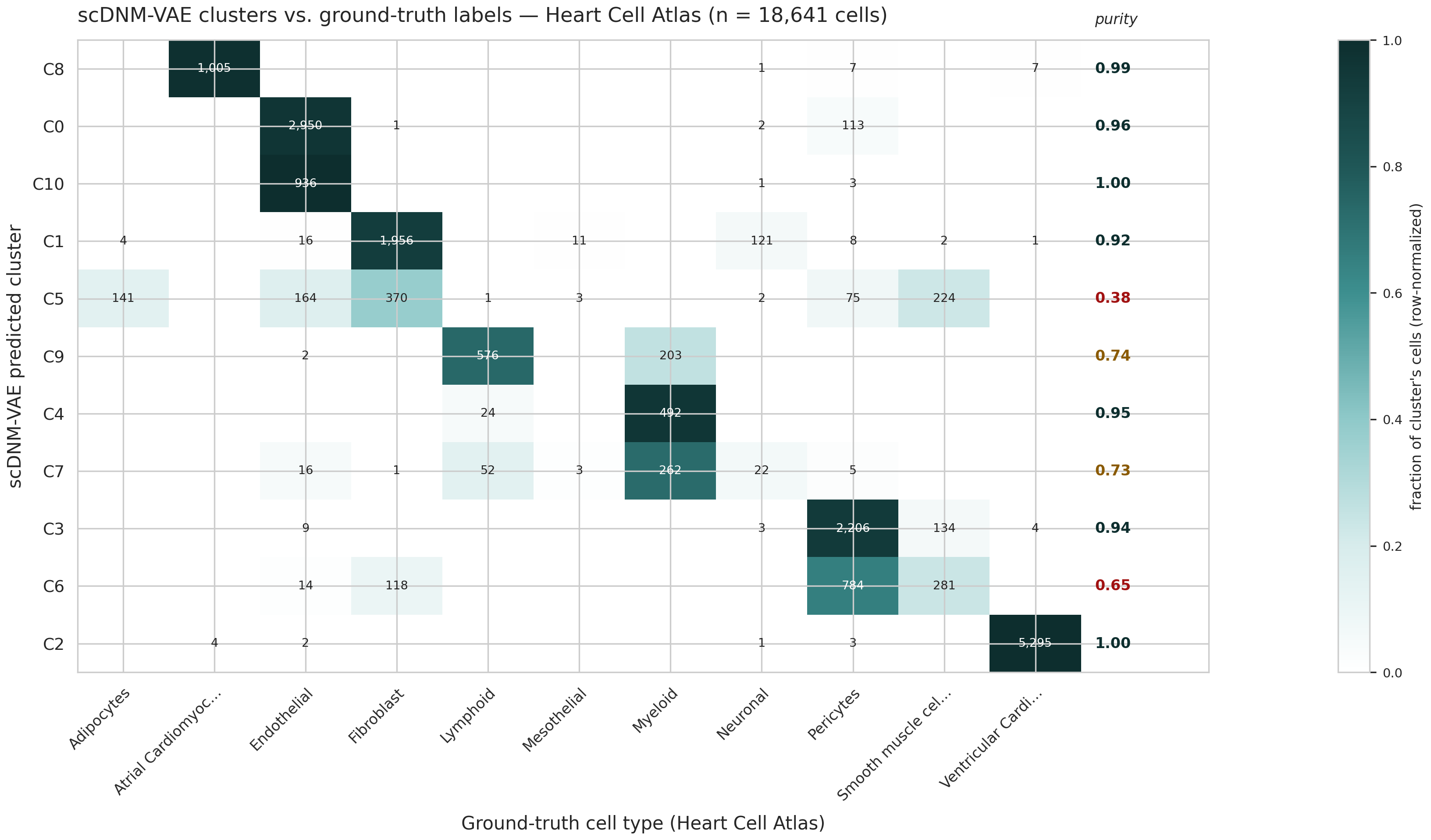}
\caption{\textbf{scDNM-VAE clusters versus ground-truth labels on the Human Heart Cell Atlas.} Row-normalized confusion matrix for the best-ARI Heart run in the five-seed benchmark ($n=18{,}641$ cells; seed 42). Rows are populated predicted clusters and columns are expert-curated cell types. Cell colour gives the within-cluster fraction and each nonzero cell is annotated with its raw count. The bars at right report per-cluster purity, defined as the largest within-row fraction. Rows are ordered by their dominant reference type. Ventricular and atrial cardiomyocytes are recovered predominantly in C0 and C1, respectively; endothelial cells are divided between C5 and C8, and lymphoid cells between C3 and C9. C10 contains only 11 cells and should not be interpreted as a stable biological population. Reference labels were used only for this external evaluation and not for training or cluster assignment.}
\label{fig:heartconf}
\end{figure}

The best-ARI Heart run from the five-seed benchmark contains ten populated clusters (Figure~\ref{fig:heartconf}). Nine of these clusters have purity between 0.83 and 0.99 against the expert-curated cell-type labels; the remaining cluster, C10, contains only 11 cells and has purity 0.45. The two cardiomyocyte compartments are recovered separately: C0 contains 5{,}301 ventricular cardiomyocytes among 5{,}455 cells (purity 0.97), whereas C1 contains 954 atrial cardiomyocytes among 1{,}043 cells (purity 0.91).

Several abundant lineages are divided into label-consistent subclusters. Endothelial cells dominate both C5 (2{,}908 of 3{,}171 cells; purity 0.92) and C8 (1{,}163 of 1{,}179; purity 0.99). Lymphoid cells similarly dominate C3 (206 of 211; purity 0.98) and C9 (405 of 444; purity 0.91). C4 is predominantly myeloid (849 of 895; purity 0.95), C6 is predominantly fibroblast (2{,}433 of 2{,}706; purity 0.90), and C7 is predominantly pericytic (2{,}944 of 3{,}526; purity 0.83), with most of the residual C7 cells annotated as smooth muscle (562 cells). Thus, the principal departures from a one-cluster--one-label correspondence are splitting within endothelial and lymphoid compartments and partial mixing along the pericyte--smooth-muscle boundary. Whether these splits represent biologically meaningful substructure or redundant partitioning requires differential-expression or pathway-level comparison and cannot be established from the confusion matrix alone.
 
\subsection{Architectural interpretability through signed synaptic gating}

Unlike post-hoc explanation methods, scDNM-VAE exposes the parameters of the clustering rule itself. For each cluster $k$ and latent dimension $d$, the gate
\[
A_{kd}(z_d)=\sigma\!\left(k_{\mathrm{sharp}}(z_dW_{kd}-Q_{kd})\right)
\]
is determined by a learned weight $W_{kd}$ and threshold $Q_{kd}$. Its derivative,
\[
\frac{\partial A_{kd}}{\partial z_d}
=
k_{\mathrm{sharp}}W_{kd}A_{kd}(1-A_{kd}),
\]
shows that the sign of $W_{kd}$ determines whether the gate increases or decreases with the latent coordinate. The magnitude $|W_{kd}|$ controls the steepness of this dependence, and, for $W_{kd}\neq0$, the transition midpoint occurs at $z_d=Q_{kd}/W_{kd}$. The complete matrices $\Wsyn$ and $\Qsyn$ therefore provide a compact, model-internal description of how every cluster reads every latent dimension, without fitting a separate post-hoc explanation model.

Because the VAE latent coordinates are not confined to a fixed input
interval, and the empirical range of each latent dimension varies across
datasets and random seeds, we interpret the trained gates continuously
through their signed weights, threshold parameters, and realized
activation functions rather than assigning them to discrete excitatory,
inhibitory, constant-0, or constant-1 categories. The ablation analysis below complements this parameter-level access by
testing whether the latent dimensions receiving the largest absolute
synaptic weights measurably influence cluster assignments.

\subsection{Faithfulness: how much of the clustering decision do the highest-weight synaptic dimensions carry?}

A parameter-level interpretation is useful only if the dimensions it highlights influence the model's decisions rather than merely co-varying with them. Because the signed weight--threshold functions are part of the trained
clustering rule, we can directly test whether a simple magnitude ranking
faithfully summarizes which latent dimensions influence assignments. We therefore tested the quantity the clustering head weights most heavily: each cluster's highest-magnitude synaptic dimensions. For each cluster we ranked all ten latent dimensions by synaptic-weight magnitude $|W_{kd}|$ and ablated the top three by neutralizing their input to that cluster's gates ($z_d W_{kd}\!\rightarrow\!0$, so the gate depends only on its threshold), recomputed the soft assignments, and measured the fraction of the cluster's cells that were reassigned (the defection rate). We compared this against two controls: ablating the top dimensions of a different, randomly chosen cluster, and ablating three random dimensions (each averaged over 20 draws).

\begin{figure}[H]
\centering
\includegraphics[width=0.8\textwidth]{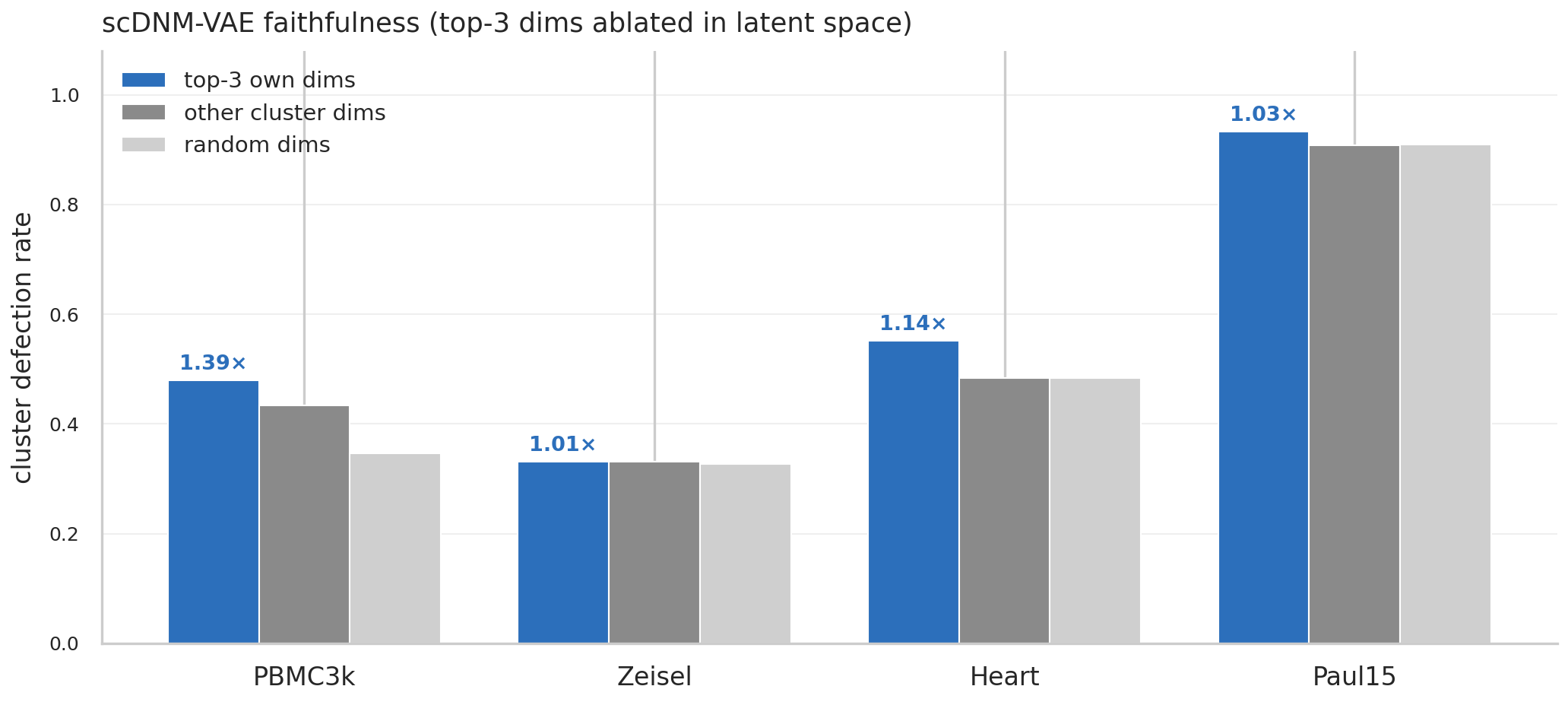}
\caption{\textbf{Highest-magnitude synaptic-weight ablation.} Mean populated-cluster defection rate after ablating a cluster's own three highest-$|W_{kd}|$ latent dimensions (blue), the top three dimensions of another populated cluster (dark grey), or three random dimensions (light grey). Controls are averaged over 20 draws for the seed-42 model of each dataset. Own top-3 ablation produces the largest mean defection rate on all four datasets, but its advantage over the random control is modest ($1.03$--$1.44\times$). Paul15 is highly sensitive to all three perturbations, consistent with a decision distributed broadly across its latent representation.}
\label{fig:faith}
\end{figure}

On all four datasets, ablating a cluster's own top synaptic dimensions caused more defection than either control (Table~\ref{tab:faith}, Figure~\ref{fig:faith}). The margin over random dimensions remains modest and dataset dependent: $1.44\times$ on PBMC3k (25.9\% vs.\ 17.9\%), $1.10\times$ on Zeisel (20.2\% vs.\ 18.4\%), $1.23\times$ on Heart (14.2\% vs.\ 11.5\%), and $1.03\times$ on Paul15 (84.6\% vs.\ 82.3\%). The clearest relative separation is therefore observed on PBMC3k, whereas the Paul15 perturbations are nearly indistinguishable despite their high absolute defection rates. The other-cluster control lies close to the random control on every dataset.

The absolute defection rates are as informative as the ratios: random three-dimensional ablation reassigns 18\% of cells on PBMC3k but 82\% on Paul15, showing that perturbation sensitivity is strongly dataset dependent. These results are consistent with a clustering decision distributed across the ten-dimensional latent space rather than concentrated in a sparse subset of dimensions, and the top-$|W_{kd}|$ dimensions are a consistent but not dominant contributor. We therefore do not claim that these dimensions are the primary causal drivers of cluster membership; the ablation supports the weaker statement that they carry decision-relevant signal on each dataset tested while the remaining dimensions carry a comparable amount. This limits the specificity of a simple magnitude ranking, while the complete signed weight-threshold decision function remains directly inspectable. Thus, the negative result concerns
the adequacy of magnitude ranking as a summary, not the accessibility of
the complete signed weight--threshold decision function itself.

\begin{table}[t]
\centering
\caption{\textbf{Faithfulness of the highest-magnitude synaptic dimensions by ablation.} Mean cluster defection rate (fraction of a cluster's cells reassigned) after ablating, per populated cluster, its own top-3 synaptic dimensions versus two controls: the top dimensions of a different cluster and three random dimensions. Both controls are averaged over 20 draws using the seed-42 model for each dataset. Ablation neutralizes the dimension's input to that cluster's gates ($z_d W_{kd}\!\rightarrow\!0$; Methods). Own-dimension ablation exceeds the random control on every dataset by $1.03$--$1.44\times$, while the high control rates on Paul15 indicate broadly distributed perturbation sensitivity. This faithfulness ablation is distinct from the architectural MLP-DEC comparison summarized across five seeds in Table~\ref{tab:ari}.}
\label{tab:faith}
\small
\begin{tabular}{lccc}
\toprule
\textbf{Dataset} & \textbf{Own top-3 dims} & \textbf{Other cluster's dims} & \textbf{Random dims} \\
\midrule
PBMC3k & \textbf{0.259} & 0.191 & 0.179 \\
Zeisel & \textbf{0.202} & 0.184 & 0.184 \\
Heart  & \textbf{0.142} & 0.121 & 0.115 \\
Paul15 & \textbf{0.846} & 0.824 & 0.823 \\
\bottomrule
\end{tabular}
\end{table}

\subsection{Gene-module decomposition (Stage 2)}

The signed-gating analysis above explains how latent dimensions contribute
to the clustering rule, but it does not indicate which gene programs those
dimensions represent. To provide a gene-level view of the learned clusters,
Stage~2 freezes the encoder and passes the latent representations through a
20-branch dendritic gene-module decoder. One branch is aligned with the
differential-expression signature of each populated Stage-1 cluster, while
the remaining branches capture background genes. Cluster-specific signatures
are obtained by one-vs-rest Wilcoxon ranking of the highly variable genes,
with mitochondrial genes excluded to reduce metabolic and technical effects.
The decoder is then trained with a reconstruction objective together with an
activation term that encourages each module to be active in its corresponding
Stage-1 cluster. Importantly, this supervision uses the predicted Stage-1
cluster assignments rather than ground-truth cell-type labels. Because the
encoder and Stage-1 assignments remain fixed, Stage~2 does not alter the
clustering solution; its role is to characterize the learned clusters at the
gene-program level.

This module alignment should therefore be interpreted as a characterization
of the learned partition rather than as independent biological validation.
In the selected best-seed runs, the recovered signatures nevertheless contain
well-established lineage markers. In PBMC3k, B-cell modules include
\emph{CD79A}, \emph{CD74}, \emph{CD79B}, and \emph{MS4A1}, whereas
CD14-monocyte modules include \emph{LYZ}, \emph{TYROBP}, \emph{LST1},
and \emph{S100A9}. Zeisel modules recover \emph{Gad1}/\emph{Gad2} for
interneurons and \emph{Plp1}/\emph{Mbp}/\emph{Mobp}/\emph{Mog} for
oligodendrocytes. In Heart, ventricular and atrial cardiomyocyte clusters are
associated with \emph{RYR2}/\emph{FHL2}/\emph{SORBS2} and
\emph{MYL7}/\emph{MYH6}/\emph{NPPA}, respectively. Paul15 shows coherent
erythroid and megakaryocytic signatures, but several myeloid progenitor
clusters share overlapping programs, consistent with the weaker
cluster--label agreement observed for this dataset.
\section{Discussion}

In this study, we proposed scDNM-VAE, a two-stage model for deep clustering of single-cell RNA-sequencing data that computes cluster membership using learned signed weights and thresholds. Comparisons across four benchmark datasets with distinct biological structures showed that scDNM-VAE provided a directly inspectable decision function while ranking among the two best-performing methods on every dataset. These findings demonstrate that making the cluster-assignment mechanism architecturally interpretable does not necessarily entail a loss of accuracy and may provide a performance advantage on some datasets.

The variation in scDNM-VAE performance across datasets suggests that the effectiveness of dendritic gating depends on the geometry of cell populations in the latent space. Because this mechanism models the contribution of each latent coordinate to cluster membership through a gate with a defined direction and transition point, it can form effective decision boundaries when cell types separate along particular latent axes. The superior performance of the dendritic head over the matched DEC head on PBMC3k and Zeisel is consistent with this interpretation. In contrast, the DEC head performed better on Heart and Paul15. In Paul15,
where differentiation follows a continuous biological trajectory, the
absence of sharp boundaries between neighboring cellular states may have
favored distance-based Student-$t$ assignment. Thus, the performance difference between the dendritic and centroid-based heads appears to reflect the compatibility between their decision geometries and the latent structure of the data rather than the general superiority of either approach.

Nevertheless, differences in performance across datasets cannot be attributed solely to the biological structure of the cell populations. Cell number, class imbalance, the granularity of the reference labels, measurement noise, and the amount of signal retained after preprocessing may also have influenced the results. In Paul15, in particular, defining closely related progenitor states as separate reference classes may have caused biologically gradual transitions to be evaluated as discrete clusters, thereby lowering the ARI values of all methods. Evaluating the methods using the same data partitions, latent dimensionality, preprocessing steps, and random seeds reduced some of these potential confounding effects. Moreover, the matched ablation analysis, in which only the clustering head was changed, supports the conclusion that at least part of the observed differences arose from the cluster-assignment mechanism itself rather than from the encoder or training protocol.

The perturbation analysis provides a more detailed view of the decision mechanism implemented by the dendritic gates. Ablating the latent dimensions with the largest absolute weights produced more cluster reassignments than ablating randomly selected dimensions across all datasets, supporting the conclusion that the learned weights contain decision-relevant information. However, the limited difference between the two ablation strategies, together with the high reassignment rate produced by random perturbation in Paul15, indicates that cluster decisions are not determined by only a few dominant latent dimensions. Weight magnitudes can therefore be used to identify decision-relevant contributions, but they should not be interpreted as independently decisive features. The inspectability of scDNM-VAE rests on access to the complete decision function defined by its signed weights and thresholds rather than on the ranking of a small number of weights.

scDNM-VAE shares the general objective of incorporating
interpretability into the model architecture with previous
single-cell approaches, including expiMap~\cite{32},
VEGA~\cite{33}, scETM~\cite{35}, and LDVAE~\cite{34},
but addresses a different question. Whereas these methods
primarily explain the biological meaning of latent factors
or gene programs, scDNM-VAE makes the decision rule by
which cells are assigned to clusters directly inspectable.
Our approach therefore complements existing methods by
extending interpretability to the cluster-assignment
mechanism, an aspect that has received comparatively
less attention in single-cell analysis.

Complementary approaches to molecular interpretation have
integrated feature selection with external biological evidence.
For example, the SAFE framework combines feature selection,
PubMed-based literature mining, and optional pathway
enrichment to support the interpretation of candidate
molecular features~\cite{47}. Such evidence-based analyses
could complement the gene-module characterization provided
by scDNM-VAE, although they were not integrated into the
present model.

The comparative results further indicate that the advantages of different
model assumptions depend on the structure of the data. The small advantage
of scVI+KMeans on Zeisel may be related to its count-aware probabilistic
modeling~\cite{1}. MLP-DEC uses the centroid-based Student-$t$ assignment
introduced in DEC~\cite{4}; its advantage on Paul15 may reflect the
suitability of this assignment mechanism for continuous hematopoietic
trajectories. Conversely, MLP-DEC produced more compact clusters but lower
ARI than scDNM-VAE on three datasets, showing that latent-space compactness
does not necessarily imply stronger agreement with biological annotations.
These findings position scDNM-VAE as a complementary approach that combines
competitive clustering performance with a directly inspectable
cluster-assignment mechanism.

The principal methodological implication of these findings is that interpretability in deep clustering models need not be provided solely through post-hoc explanation methods. The cluster-assignment mechanism of scDNM-VAE, which consists of signed weights and thresholds, allows researchers to directly examine the direction and transition points of latent dimensions involved in assigning a cell to a particular cluster. This property may make the method a useful alternative in single-cell studies where auditing cluster decisions and translating them into biological hypotheses are important. However, given the differences in performance across datasets, scDNM-VAE should not be regarded as a universal replacement for existing clustering methods, but rather as a complementary tool to be evaluated alongside standard approaches when interpretability is a priority.

Several limitations constrain the interpretation of the performance results. First, scDNM-VAE assumes that the number of clusters is known in advance; because this information is generally unavailable in real-world applications, the performance reported here may be optimistic. Second, applying a Gaussian reconstruction loss to log-normalized data may not fully capture sampling variability and overdispersion in raw counts, potentially reducing the performance of scDNM-VAE relative to count-aware methods, particularly in sparse and noisy datasets. In addition, evaluating the model on four datasets and against a limited number of comparators restricts the generalizability of the observed advantages to other tissues, experimental platforms, and dataset sizes. Reference cell labels also do not represent error-free biological ground truth, and ARI may therefore underestimate biologically meaningful structures, particularly in datasets such as Paul15 that contain continuous differentiation trajectories.

The interpretability results should be considered with similar caution. The seed-to-seed variability observed on Zeisel suggests sensitivity to initialization and indicates that interpretations based on a single best-performing run may overstate performance. Furthermore, the weight-ablation analysis was conducted using only one seed, and it therefore remains unclear whether the influence of high-weight latent dimensions is preserved across different initializations. Because the gene modules obtained in Stage 2 were learned using the clusters generated in Stage 1, they do not constitute independent biological validation and may propagate clustering errors from the first stage. The signed weights, thresholds, and gene modules should therefore be interpreted as model-based explanations requiring experimental or external validation rather than as validated causal biological mechanisms.

\section{Conclusions}

In this study, we proposed scDNM-VAE, a two-stage framework that replaces centroid-based cluster assignment with signed weight--threshold gates whose direction, steepness, and transition location can be inspected directly. Across four scRNA-seq datasets, scDNM-VAE achieved the highest mean ARI on PBMC3k and the Human Heart Cell Atlas, performed comparably to the best-performing method on Zeisel, and remained below MLP-DEC on Paul15. The matched head-only ablation further showed that the dendritic head performed better on PBMC3k and Zeisel, whereas the Student-$t$ DEC head achieved better results on Heart and Paul15. These findings indicate that the effect of the dendritic assignment mechanism depends on the structure of the dataset rather than reflecting uniform superiority.

Ablating the highest-weight synaptic dimensions affected cluster assignments more than random ablation, although the modest differences indicated that decision-relevant information was distributed across the latent space. Moreover, freezing the clustering solution before Stage~2 enabled the extraction of cluster-associated gene signatures without altering the learned assignments. Overall, signed dendritic gating provides a directly inspectable alternative for single-cell clustering while maintaining competitive performance across the datasets evaluated here. Future studies should evaluate the method in larger independent datasets from different tissues and experimental platforms and test the stability of the dendritic gates using designs in which the number of clusters is inferred from the data, raw counts are modeled directly, and interpretability analyses are repeated across multiple random initializations. In particular, comparisons between discrete cell populations and continuous cellular trajectories will help determine the data structures in which dendritic gating provides an advantage over centroid-based assignment.

\section{Methods}

\subsection{Datasets}
PBMC3k: peripheral blood mononuclear cells ($n=2{,}638$, 8 cell types) from 10x Genomics. Zeisel Cortex: mouse somatosensory cortex (S1) and hippocampal CA1 region ($n=3{,}005$ cells)~\cite{8}; benchmarking here uses the dataset's standard seven-category reference-label taxonomy, coarser than the nine major cell classes reported in the original study. Human Heart Cell Atlas: a fixed random subsample of the adult human heart atlas of Litvi\v{n}ukov\'a et al.~\cite{7}, distributed with scvi-tools~\cite{43} ($n=18{,}641$ cells, 11 annotated cell types); we used this subsample as provided, together with the \texttt{cell\_type} annotations it carries as ground truth, rather than subsampling the $\sim$486K-cell atlas ourselves. Paul15: mouse hematopoietic progenitors ($n=2{,}730$, 19 cell types)~\cite{9}. Exact loaders and accessions for all four benchmarks are listed under Data availability.

\subsection{Preprocessing}
Filtered cells $<200$ genes and genes $<3$ cells. Cells with more than 20\% of counts mapping to mitochondrial genes were removed as putative dead or damaged cells; mitochondrial genes themselves were retained as biologically informative (e.g., for high-energy cardiac populations). Across all four benchmarks no cell exceeded this threshold (maximum mitochondrial fraction 19.95\% on the Human Heart Cell Atlas), consistent with the published quality control of these datasets. Data were normalized to 10{,}000 counts/cell, $\log(1+x)$-transformed, reduced to the top 2{,}000 highly variable genes (Seurat method), and scaled to unit variance with max~$=10$. The identical mitochondrial filter was applied to the scDNM-VAE, scVI, and annotation data paths. For scVI baselines, raw counts were preserved with the ZINB likelihood~\cite{1}.

\subsection{Dendritic Neuron Model (DNM)}
The DNM~\cite{2} is a biologically inspired architecture that mimics how a single neuron integrates inputs along its dendritic tree through four sequential layers---synaptic, dendritic, membrane, and soma. In the \textbf{synaptic layer}, each input $x_i$ to dendrite $j$ passes through a sigmoid gate $y_{ij}=\sigma\!\big(k(w_{ij}x_i-q_{ij})\big)$. The derivative $\partial y_{ij}/\partial x_i=k w_{ij}y_{ij}(1-y_{ij})$ makes the gate's directional dependence explicit: positive weights produce increasing gates, negative weights produce decreasing gates, $|w_{ij}|$ controls transition steepness, and $q_{ij}$ shifts the transition location. The \textbf{dendritic layer} combines gated outputs along each branch multiplicatively, $d_j=\prod_i y_{ij}$; the \textbf{membrane layer} aggregates branch outputs by summation, $u=\sum_j d_j$; and the \textbf{soma layer} applies a final sigmoid, $o=\sigma(u)$. This threshold-gated structure makes the role of each connection directly inspectable through its trained weight, threshold, and activation function. scDNM-VAE adapts this principle to clustering, using signed per-dimension gates in the Stage-1 clustering head and geometric-mean branch aggregation in the Stage-2 gene-module decoder.

\subsection{scDNM-VAE architecture}
\textbf{Architecture.} The variational encoder is a two-layer MLP (input, 512 and 128) with BatchNorm, ReLU activations, and 10\% input dropout, projecting onto a 10-dimensional latent space. We parameterize the variational posterior with a mean vector $\boldsymbol{\mu}$ and a log-variance vector $\log\boldsymbol{\sigma}^2$ (clamped to $[-4,4]$ for numerical stability), and draw latent samples via the standard reparameterization trick $\mathbf{z}=\boldsymbol{\mu}+\boldsymbol{\varepsilon}\cdot\boldsymbol{\sigma}$ with $\boldsymbol{\varepsilon}\sim\mathcal{N}(\mathbf{0},\mathbf{I})$. The decoder mirrors the encoder symmetrically. The dendritic clustering head computes a sigmoid gate per (cluster, latent dimension) pair,
\[
A_{kd}=\sigma\!\left(k_{\mathrm{sharp}}(z_dW_{kd}-Q_{kd})\right),
\]
with learnable thresholds capped above by $Q_{\max}=1.5$. During Phase~2, gating sharpness follows
\[
k_{\mathrm{sharp}}(t)=1.5+0.5\min\!\left(\frac{t}{0.25T_2},1\right),
\]
where $t$ is the Phase~2 epoch and $T_2=300$; it remains fixed at 2.0 during Phase~3. Its derivative is
\[
\frac{\partial A_{kd}}{\partial z_d}=k_{\mathrm{sharp}}W_{kd}A_{kd}(1-A_{kd}),
\]
so $W_{kd}>0$ defines an increasing gate and $W_{kd}<0$ a decreasing gate; $|W_{kd}|$ controls transition steepness, and for $W_{kd}\neq0$ the transition midpoint is $Q_{kd}/W_{kd}$. Because the VAE latent coordinates are not restricted to the bounded input interval assumed by classical four-state DNM partitions, we do not discretize the trained gates into excitatory, inhibitory, constant-0, or constant-1 categories. Instead, architectural interpretation is based on the continuous signed weight-threshold functions actually used to compute cluster assignments. No final threshold lay at the imposed upper boundary; the maximum final $Q_{kd}$ value across the four datasets was $1.21$.

\textbf{Training objective.} The full loss combines reconstruction, DEC-style clustering, variational KL regularization, and an unsupervised occupancy-prior term:
\begin{equation}
\mathcal{L}=\Lrecon+\gamma\,\Lcluster+\beta\,\Lkl+\lambda_{\mathrm{occ}}\Locc,
\end{equation}

 with $\gamma=2.0$ and dataset-specific fixed KL coefficients:
$\beta=10^{-3}$ for PBMC3k and Paul15 and
$\beta=5\times10^{-4}$ for Zeisel and Heart. The DEC target distribution $p_{ik}\propto q_{ik}^2/\sum_i q_{ik}$ is recomputed every 5 epochs from the current soft assignments $q_{ik}$. Let $r_{ik}$ denote the soft assignment from the KMeans initialization and define $\pi_k^{\mathrm{KM}}=N^{-1}\sum_i r_{ik}$ and $\bar q_k=N^{-1}\sum_i q_{ik}$. The occupancy loss is
\begin{equation}
\Locc=D_{\mathrm{KL}}\!\left(\boldsymbol{\pi}^{\mathrm{KM}}\,\middle\|\,\bar{\mathbf q}\right)
=\sum_k \pi_k^{\mathrm{KM}}\log\frac{\pi_k^{\mathrm{KM}}+\epsilon}{\bar q_k+\epsilon},
\end{equation}
with $\lambda_{\mathrm{occ}}=1.0$ and $\epsilon=10^{-8}$ for numerical stability. This term uses no reference labels: it preserves the aggregate occupancy profile of the populated initialization while allowing individual cells and cluster boundaries to change. It is active only during the joint optimization phases.

\subsection{KL-regularization settings}

For the reported experiments, the KL coefficient was set to
$\beta=10^{-3}$ for PBMC3k and Paul15 and
$\beta=5\times10^{-4}$ for Zeisel and Heart. Within each dataset,
the same value was held fixed across all five random seeds. These values
are reported as dataset-specific implementation settings for the benchmark
experiments and should not be interpreted as estimates of dataset noise or
as a generally applicable data-driven rule for selecting $\beta$.

A systematic sensitivity analysis of the KL coefficient was not performed;
the reported values should therefore not be interpreted as generally optimal
for these datasets.

\subsection{Training protocol}
scDNM-VAE and the MLP-DEC ablation are optimized with Adam (weight decay $10^{-5}$) using full-batch updates and a cosine-annealing learning-rate schedule; the scVI baseline follows its own training protocol, given in Section~\ref{sec:scvi}. Training proceeds in three phases. In Phase~1 (300 epochs, learning rate $10^{-3}$), the encoder and decoder are pretrained on $\Lrecon+\beta\Lkl$ (clustering and occupancy terms switched off), allowing the latent space to organize before any clustering signal is introduced. The KL term was included in Phase~1 for all four benchmarks because all reported $\beta$ values were positive. In Phase~2 (300 epochs, learning rate $5\times10^{-4}$), KMeans ($n_{\mathrm{init}}=20$, with \texttt{random\_state} set to the corresponding training seed) is fitted to the posterior-mean representations $\boldsymbol{\mu}$ produced by the pretrained encoder. The resulting $K$ centre vectors are copied into the rows of $\Wsyn$, and their soft assignments define $\boldsymbol{\pi}^{\mathrm{KM}}$. For each cluster $k$, let $\delta_k$ denote the Euclidean distance between its row $\mathbf{W}_k$ and the nearest other row of $\Wsyn$; every threshold in that cluster is initialized as
\[
Q_{kd}=\operatorname{clip}\!\left(\frac{\delta_k}{3},\,0.05,\,0.8Q_{\max}\right),
\qquad d=1,\ldots,D,
\]
with $Q_{\max}=1.5$, after which the full objective including $\Locc$ is optimized end-to-end. During the first 75 Phase~2 epochs, $k_{\mathrm{sharp}}$ increases linearly from 1.5 to 2.0 and then remains fixed. In Phase~3 (50 epochs, learning rate $2\times10^{-4}$), the same threshold-initialization rule is applied again using the optimized $\Wsyn$, and the model is briefly fine-tuned with $k_{\mathrm{sharp}}=2.0$ and the occupancy term retained. Each model is trained with five random seeds (42, 7, 123, 256, 999); mean $\pm$ sample standard deviation is the primary report and the best seed is secondary.
\paragraph{Selection of representative runs}
The primary benchmark uses all five seeds. The paired UMAP panels and Heart confusion matrix use the model with the highest unrounded ARI within each dataset, as stated in their captions. The faithfulness analysis instead uses seed 42 for every dataset to avoid selecting an interpretability result on the basis of its effect size.

\paragraph{Determinism and hardware}
The benchmark values in Tables~\ref{tab:ari} and~\ref{tab:bench} were
obtained from controlled local CPU runs. Python, NumPy, and PyTorch
random-number generators were reset for every seed. scDNM-VAE and MLP-DEC
used identical deterministic implementations of KMeans and evaluation
metrics, while scVI was run with scvi-tools~1.4.2 and its
method-appropriate raw-count input. The reproducibility package records
the software versions, training seeds, model parameters, and cluster
assignments used for the reported analyses. It will be made publicly
available upon publication and is available from the corresponding author
upon reasonable request.
\subsection{Matched head-only VAE--DEC ablation}
\label{sec:headonly}

To isolate the clustering head while preserving the remainder of the
model, we constructed a matched VAE--DEC variant in which only the signed
dendritic assignment function was replaced by a standard Student-$t$
centroid head. The variant retained the stochastic scDNM-VAE encoder,
the mirrored decoder, the 10-dimensional latent representation, the
Gaussian reparameterization step, the reconstruction objective, and the
dataset-specific KL coefficient. It also used the same normalized
highly-variable-gene matrices, five random seeds (42, 7, 123, 256, and
999), deterministic KMeans implementation, and number of clusters as
scDNM-VAE.

Phase~1 pretrained the VAE for 300 epochs at a learning rate of
$10^{-3}$. KMeans was then fitted to the posterior means, and its centres
initialized the learnable DEC centroids. The corresponding soft
assignments defined the same KMeans-derived non-uniform occupancy prior
used by scDNM-VAE. Phase~2 jointly optimized reconstruction, KL,
DEC clustering, and occupancy losses for 300 epochs at
$5\times10^{-4}$. The DEC soft assignment was

\begin{equation}
q_{ik}^{\mathrm{DEC}} =
\frac{
\left(1+\lVert\boldsymbol{\mu}_i-\mathbf{c}_k\rVert^2\right)^{-1}
}{
\sum_{k'}
\left(1+\lVert\boldsymbol{\mu}_i-\mathbf{c}_{k'}\rVert^2\right)^{-1}
},
\end{equation}

where $\mathbf{c}_k$ is the learnable centroid of cluster $k$. The usual
DEC target distribution was recomputed every five epochs. To match the
duration and lower learning rate of scDNM-VAE's third phase, optimization
continued for 50 additional epochs at $2\times10^{-4}$. No threshold
reset was performed because the DEC head has no threshold parameters.
The final assignments were obtained by
$\arg\max_k q_{ik}^{\mathrm{DEC}}$.
\subsection{MLP-DEC ablation} \label{sec:mlpdec}
MLP-DEC is a broader architectural comparator rather than the matched
head-only ablation described in Section~\ref{sec:headonly}. It uses a
deterministic encoder and a standard DEC head while retaining the decoder,
optimizer, and first two training phases of scDNM-VAE; it has no KL term,
occupancy regularization, or analogue of the threshold-refinement phase.

\textbf{Encoder and decoder.} The encoder uses the same two-hidden-layer MLP (input$\rightarrow$512$\rightarrow$128) with BatchNorm, ReLU, and 10\% input dropout as scDNM-VAE, but maps \emph{deterministically} to the 10-dimensional latent space through a final linear layer, with no variational posterior and hence no KL term ($\mathbf{z}=f_{\mathrm{enc}}(\mathbf{x})$). The decoder mirrors it symmetrically (128, 512 hidden units), identical to scDNM-VAE's.

\textbf{Clustering head.} In place of the synaptic weights and thresholds, the head holds $K$ learnable centroids $\{\boldsymbol{\mu}_k\}$ in the latent space. The soft assignment of cell $i$ to cluster $k$ is a Student-$t$ kernel with one degree of freedom,
\begin{equation}
q_{ik}=\frac{\bigl(1+\lVert \mathbf{z}_i-\boldsymbol{\mu}_k\rVert^2\bigr)^{-1}}{\sum_{k'}\bigl(1+\lVert \mathbf{z}_i-\boldsymbol{\mu}_{k'}\rVert^2\bigr)^{-1}},
\end{equation}
and the auxiliary target distribution sharpens these assignments,
\begin{equation}
p_{ik}=\frac{q_{ik}^2/f_k}{\sum_{k'} q_{ik'}^2/f_{k'}},\qquad f_k=\sum_i q_{ik},
\end{equation}
recomputed every five epochs. The clustering loss is $\Lcluster=\mathrm{KL}(P\,\Vert\,Q)=\sum_{i,k} p_{ik}\log(p_{ik}/q_{ik})$---the \emph{same} auxiliary-distribution loss optimized by scDNM-VAE's head; only the assignment $q_{ik}$ differs, using Euclidean distance to centroids here versus the dendritic gate score $\mathrm{dist}_k=\sum_d(1-A_{kd})^2$ in scDNM-VAE.

\textbf{Training.} The ablation follows the same pretraining-then-joint schedule as scDNM-VAE: a 300-epoch reconstruction-only pretraining (Adam, learning rate $10^{-3}$), KMeans initialization of the centroids ($n_{\mathrm{init}}=20$) on the pretrained latent space, and a 300-epoch joint phase (learning rate $5\times10^{-4}$) minimizing $\mathcal{L}=\Lrecon+\gamma\,\Lcluster$ with $\gamma=2$. It does \emph{not} include scDNM-VAE's 50-epoch Phase 3, which re-initializes the thresholds $\Qsyn$ from the stabilized $\Wsyn$: the DEC head has no threshold parameter and therefore no analogue of that refinement.

\textbf{Scope of the ablation.} The two models thus share the encoder trunk, decoder, reconstruction loss, clustering loss $\Lcluster$, optimizer, and the first two training phases, and differ in (i)~the clustering head---$K$ threshold-gated synaptic weight/threshold pairs ($\Wsyn,\Qsyn$; $2\times K\times 10$ parameters) versus $K$ Euclidean centroids ($K\times 10$ parameters); (ii)~the variational versus deterministic encoding of the latent space; and (iii)~the threshold-refinement Phase 3. Because these components differ together, the performance gap in
Table~\ref{tab:ari} is attributable to the combined architectural
replacement rather than to dendritic gating or clustering-head
replacement in isolation.

\subsection{scVI baseline}\label{sec:scvi}
As an external reference point, we benchmarked against scVI~\cite{1}, a widely used variational autoencoder for scRNA-seq data that models gene counts with a ZINB likelihood. We trained scVI with the same latent dimensionality as scDNM-VAE ($n_{\mathrm{latent}}=10$) for 150 epochs with batch size 128 and a 50-epoch KL warm-up schedule. Because scVI does not include a clustering objective, we obtained discrete cluster assignments by running KMeans ($n_{\mathrm{init}}=20$) on the resulting latent representations, with the number of clusters set to the ground-truth value for each dataset (Table~\ref{tab:bench}). To match the reporting protocol used for scDNM-VAE and the MLP-DEC ablation, scVI was trained with the same five random seeds (42, 7, 123, 256, 999); we report the best-seed ARI in Table~\ref{tab:bench} alongside the mean $\pm$ standard deviation across seeds in Table~\ref{tab:ari}.

\subsection{Faithfulness ablation of the highest-magnitude synaptic dimensions}
To test how much of each cluster assignment a cluster's highest-magnitude synaptic dimensions account for, we used a single scDNM-VAE per dataset (seed 42) and, for each cluster, ranked all ten latent dimensions by synaptic-weight magnitude $|W_{kd}|$. We then ablated the cluster's top three dimensions by neutralizing their input to that cluster's gates: the drive $z_d W_{kd}$ is set to zero for those (cluster, dimension) pairs only, so that $A_{kd}=\sigma(-k_{\mathrm{sharp}}Q_{kd})$ no longer responds to the cell's latent value while the remaining clusters are untouched. We then recomputed the soft assignments from $\mathrm{dist}_k=\sum_d(1-A_{kd})^2$ and recorded the defection rate: the fraction of cells originally assigned to the cluster that were reassigned elsewhere. Note that simply dropping the masked terms from the sum would lower $\mathrm{dist}_k$ and make the cluster more attractive, so it cannot produce defection; neutralizing the input is the ablation that tests whether the dimension informs the decision. Each cluster was ablated independently. As controls, for each cluster we ablated (i) the top three dimensions
of a different, randomly chosen cluster, and (ii) three randomly chosen
dimensions; both controls were averaged over 20 random draws. Random
three-dimensional control sets were sampled without excluding the target
cluster's top-ranked dimensions; because these sets overlap the target top
three by 0.9 dimensions on average, this comparison provides a conservative
estimate of the contrast between top-ranked and arbitrary latent dimensions.
Dataset-level values (Table~\ref{tab:faith}) are means over clusters. This faithfulness ablation is distinct from the architectural MLP-DEC
ablation used for the benchmark comparison
(Table~\ref{tab:ari}): the former perturbs a trained model's synaptic
dimensions to test explanation sensitivity, whereas the latter benchmarks
the combined architectural replacement. 

\subsection{Cluster annotation by marker overlap}
To assess the biological coherence of the clusters discovered by scDNM-VAE, we annotated each predicted cluster with a putative cell type by overlap with curated marker-gene panels. We did not use a reference-based automated annotation tool (e.g., Azimuth); instead, each cluster was matched to the cell type whose curated marker panel showed the greatest overlap with the cluster's differentially expressed genes, and all cells in a cluster received that cluster's label. For each cluster, we identified the top differentially expressed genes against all other clusters using \texttt{sc.tl.rank\_genes\_groups} (Wilcoxon rank-sum test) on the full expression matrix prior to highly-variable-gene subsetting, with dataset-specific thresholds for the number of top genes ($n_{\mathrm{top}}=200$ for PBMC3k and Heart; $n_{\mathrm{top}}=500$ for Zeisel and Paul15), the $\log_2$-fold-change cutoff ($\ge0.10$ and $\ge0.05$, respectively) and the adjusted $p$-value cutoff ($\le0.1$ and $\le0.5$, respectively). Each cluster's top DE gene list was scored against curated cell-type marker panels using a custom weighted marker-overlap score that distinguishes between primary and secondary markers. For each candidate cell type, primary markers (the first three genes in the panel, typically canonical lineage-defining genes) were assigned weight 2, while secondary markers (the remaining genes) were assigned weight 1. Writing $h_p$ and $h_s$ for the number of primary and secondary markers recovered in a cluster's top-gene list, $n_p$ and $n_s$ for the panel sizes and $T$ for the length of the top-gene list, the score is the harmonic mean of a weighted recall $(2h_p+h_s)/(2n_p+n_s)$ and an unweighted precision $(h_p+h_s)/T$, plus an additive hit-count term $0.03\,(h_p+h_s)$. The cluster was annotated with the cell type whose markers achieved the highest marker-overlap score, subject to a minimum of two marker hits for PBMC3k and Heart and one for Zeisel and Paul15. Because the reference annotations and the panel vocabulary do not coincide, reference labels were mapped onto panel categories before scoring and cluster majorities were computed after this mapping; the complete mapping is given in Supplementary Table~S3. For PBMC3k, marker-positivity symbols were harmonized explicitly:
``CD14+ Monocytes'' and ``FCGR3A+ Monocytes'' were mapped to the
panel categories ``CD14 Monocytes'' and ``FCGR3A Monocytes'',
respectively, before exact-match scoring. Annotation agreement was evaluated among clusters meeting the prespecified $\geq 40\%$ dominant-label eligibility criterion.

\subsection{Complementary gene-module decoder (Stage 2)}
To translate the abstract latent dimensions of Stage~1 into biologically interpretable gene programs, we trained a separate dendritic decoder (DNMDecoder) on top of the frozen Stage~1 encoder. Freezing the encoder guarantees that Stage~2 cannot alter the clustering solution and serves only as a supervised, cluster-aligned gene-level interpretability layer. One branch was assigned to each populated Stage-1 cluster and masked to reconstruct that cluster's differential-expression signature genes, obtained by one-vs-rest Wilcoxon ranking (\texttt{sc.tl.rank\_genes\_groups}) on the highly-variable genes; genes not up-regulated in any cluster were pooled into background branches, and mitochondrial genes were excluded from the reconstruction target so that modules report cell identity rather than mitochondrial content. The DNMDecoder routes the frozen latent vector $\mathbf{z}$ through 20 parallel branches; each branch $m$ has its own learnable synaptic weights $\mathbf{W}_m$ and thresholds $\mathrm{Th}_m$, which apply per-dimension sigmoid gating to $\mathbf{z}$, and the branch activation $D_m$ is the geometric mean of its gate openings. The 20-dimensional branch-activation vector was LayerNorm-normalized
across branches for each cell before being used as module logits and
passed to the masked linear reconstruction layer. The DNMDecoder was trained for 500 epochs with Adam (learning rate $10^{-3}$) using a combined objective: a mean-squared-error reconstruction loss against the original (Stage~1) input plus a supervised activation term (weight 0.5) that treats the branch-activation vector as logits over modules and applies cross-entropy against each cell's cluster-assigned module (assignments taken from the Stage-1 predicted clusters, not ground-truth labels), encouraging each module to be active in its target cluster. Gradients flowed only through the DNMDecoder parameters, leaving the encoder unchanged. After training, each cluster's gene-module signature was obtained by averaging the per-cell branch activations $D_m$ across all cells assigned to that cluster.

\subsection{Evaluation metrics}
We assessed clustering performance along two complementary axes.
Label-based agreement between predicted clusters and curated reference
cell types was quantified using the adjusted Rand index (ARI) and
normalized mutual information (NMI). Both metrics were computed
separately for each random seed, and the primary benchmark reports the
mean and sample standard deviation across the five matched seeds.
Geometry-based metrics quantify separation and compactness in the learned
representation independently of external labels: silhouette score,
Calinski--Harabasz index, and Davies--Bouldin index were computed on the
latent embedding $\mathbf{z}$ (the encoder posterior mean
$\boldsymbol{\mu}$ for scDNM-VAE and the matched head-only VAE--DEC
ablation, the deterministic embedding
$\mathbf{z}=f_{\mathrm{enc}}(\mathbf{x})$ for MLP-DEC, and the latent
representation returned by \texttt{get\_latent\_representation} for
scVI). For differential-expression analysis, we used Scanpy's
\texttt{rank\_genes\_groups} function with the Wilcoxon rank-sum test and
Benjamini--Hochberg correction for multiple testing.

\section*{Data availability}
All four benchmark datasets are publicly available and were obtained programmatically. PBMC3k was loaded with \texttt{scanpy.datasets.pbmc3k()} (10x Genomics), with cell-type labels taken from the matching \path{scanpy.datasets.pbmc3k_processed()} object. Zeisel Cortex was obtained from the cortex expression matrix (\path{expression_mRNA_17-Aug-2014.loom}) mirrored for scvi-tools~\cite{43}; the primary data are those of Zeisel et al.~\cite{8}. The Human Heart Cell Atlas subsample was obtained via
\path{scvi.data.heart_cell_atlas_subsampled()}, which downloads
the fixed subsample deposited at
\url{https://ndownloader.figshare.com/files/24539942}.
The full $\sim$486K-cell/nucleus atlas from which it was drawn is
described by Litvi\v{n}ukov\'a et al.~\cite{7}. Paul15 was loaded with \texttt{scanpy.datasets.paul15()}; the primary data are those of Paul et al.~\cite{9}. Because the Heart benchmark uses this distributed subsample rather than the full atlas, the cell counts reported here (18{,}641) differ from those of the primary publication.

\section*{Code availability}
The source code implementing scDNM-VAE and the scripts used to reproduce the benchmarking and interpretability analyses will be made available in a public repository upon publication or earlier upon reasonable request for peer-review purposes.

\section*{Declaration of generative AI and AI-assisted technologies in the writing process}
During the preparation of this work the authors used ChatGPT (OpenAI) in order to
improve the readability and language of the manuscript. After using this tool, the
authors reviewed and edited the content as needed and take full responsibility for
the content of the publication.
\section*{Declaration of competing interest}
The authors declare that they have no known competing financial interests or
personal relationships that could have appeared to influence the work reported
in this paper.

\end{document}